\documentclass{ws-ijmpa}
\usepackage{graphicx}

\newcommand\rf[1]{(\ref{eq:#1})}
\newcommand\lab[1]{\label{eq:#1}}
\newcommand\nonu{\nonumber}
\newcommand\br{\begin{eqnarray}}
\newcommand\er{\end{eqnarray}}
\newcommand\be{\begin{equation}}
\newcommand\ee{\end{equation}}

\newcommand\foot[1]{\footnotemark\footnotetext{#1}}
\newcommand\lb{\lbrack}
\newcommand\rb{\rbrack}

\newcommand\llb{\left\lbrack}
\newcommand\rrb{\right\rbrack}

\newcommand\lcurl{\left\{}
\newcommand\rcurl{\right\}}
\renewcommand\({\left(}
\renewcommand\){\right)}
\newcommand\bv{\bigm\vert}               
\newcommand\bgv{\bigg\vert}              

\newcommand\bc{\begin{center}}
\newcommand\ec{\end{center}}

\newcommand\partder[2]{\frac{{\partial {#1}}}{{\partial {#2}}}}

\renewcommand\b{\beta}

\renewcommand\d{\delta}

\newcommand\eps{\epsilon}
\newcommand\vareps{\varepsilon}
\newcommand\g{\gamma}
\newcommand\G{\Gamma}

\newcommand\h{\frac{1}{2}}
\renewcommand\k{\kappa}
\renewcommand\l{\lambda}
\renewcommand\L{\Lambda}
\newcommand\m{\mu}
\newcommand\n{\nu}

\newcommand\vp{\varphi}
\renewcommand\P{\Phi}
\newcommand\pa{\partial}

\newcommand\pr{\prime}

\renewcommand\r{\rho}
\newcommand\s{\sigma}

\renewcommand\t{\tau}
\renewcommand\th{\theta}

\newcommand\wti{\widetilde}

\newcommand\cA{{\mathcal A}}

\newcommand\cE{{\mathcal E}}
\newcommand\cF{{\mathcal F}}

\newcommand\cJ{{\mathcal J}}

\newcommand\cV{{\mathcal V}}

\newcommand{\ct}[1]{\cite{#1}}
\newcommand{\bib}[1]{\bibitem{#1}}
\newcommand\PRL[3]{\textsl{Phys. Rev. Lett.} \textbf{#1}, #3 (#2)}
\newcommand\NPB[3]{\textsl{Nucl. Phys.} \textbf{B#1}, #3 (#2)}

\newcommand\PRD[3]{\textsl{Phys. Rev.} \textbf{D#1}, #3 (#2)}
\newcommand\PLA[3]{\textsl{Phys. Lett.} \textbf{#1A}, #3 (#2)}
\newcommand\PLB[3]{\textsl{Phys. Lett.} \textbf{#1B}, #3 (#2)}
\newcommand\CQG[3]{\textsl{Class. Quantum Grav.} \textbf{#1}, #3 (#2)}
\newcommand\JMP[3]{\textsl{J. Math. Phys.} \textbf{#1}, #3 (#2)}

\newcommand\AoP[3]{\textsl{Ann. of Phys.} \textbf{#1}, #3 (#2)}

\newcommand\IJMPA[3]{\textsl{Int. J. Mod. Phys.} \textbf{A#1}, #3 (#2)}
\newcommand\IJMPD[3]{\textsl{Int. J. Mod. Phys.} \textbf{D#1}, #3 (#2)}

\newcommand\JETP[3]{\textsl{Sov. Phys. JETP} \textbf{#1}, #3 (#2)}

\newcommand\Xdot{\stackrel{.}{X}}

\newcommand\rdot{\stackrel{.}{r}}

\begin{document}

\markboth{E. Guendelman, A. Kaganovich, E. Nissimov and S. Pacheva}
{Non-singular black holes from gravity-matter-brane lagrangians}

%
\catchline{}{}{}{}{}
%

\title{NON-SINGULAR BLACK HOLES FROM GRAVITY-MATTER-BRANE LAGRANGIANS}

\author{Eduardo Guendelman and Alexander Kaganovich}

\address{Department of Physics, Ben-Gurion University of the Negev\\
P.O.Box 653, IL-84105 ~Beer-Sheva, Israel \\
guendel@bgu.ac.il, alexk@bgu.ac.il}

\author{Emil Nissimov and Svetlana Pacheva}

\address{Institute for Nuclear Research and Nuclear Energy, Bulgarian Academy
of Sciences\\
Boul. Tsarigradsko Chausee 72, BG-1784 ~Sofia, Bulgaria \\
nissimov@inrne.bas.bg, svetlana@inrne.bas.bg}

\maketitle

\begin{history}
\received{}
\revised{}
\end{history}

\begin{abstract}
We consider self-consistent coupling of bulk Einstein-Maxwell-Kalb-Ramond system to
codimension-one charged {\em lightlike} $p$-brane with dynamical (variable) tension 
(\textsl{LL-brane}). The latter is described by a manifestly 
reparametrization-invariant world-volume action significantly different from
the ordinary Nambu-Goto one. We show that the \textsl{LL-brane} is the appropriate
gravitational and charge source 
in the Einstein-Maxwell-Kalb-Ramond equations of motion needed to generate a 
self-consistent solution describing {\em non-singular} black hole. 
The latter consists of de
Sitter interior region and exterior Reissner-Nordstr{\"o}m region glued
together along their common horizon (it is the inner horizon from the 
Reissner-Nordstr{\"o}m side). The matching horizon is automatically occupied 
by the \textsl{LL-brane} as a result of its world-volume lagrangian dynamics, 
which dynamically generates the cosmological constant in the interior region and
uniquely determines the mass and charge parameters of the exterior region.
Using similar techniques we construct a self-consistent {\em wormhole} solution of
Einstein-Maxwell system coupled to electrically neutral \textsl{LL-brane},
which describes two identical copies of a non-singular black hole region
being the exterior Reissner-Nordstr{\"o}m region above the inner horizon, 
glued together along their common horizon (the inner Reissner-Nordstr{\"o}m
one) occupied by the \textsl{LL-brane}. The corresponding mass and charge
parameters of the two black hole ``universes'' are explicitly determined 
by the dynamical \textsl{LL-brane} tension. This also provides an explicit
example of Misner-Wheeler ``charge without charge'' phenomenon.
Finally, this wormhole solution connecting two non-singular black holes can be 
transformed into a special case of Kantowski-Sachs bouncing cosmology solution
if instead of 
Reissner-Nordstr{\"o}m  we glue together two copies of the exterior
Reissner-Nordstr{\"o}m-de-Sitter region with big enough bare cosmological constant, 
such that the radial coordinate becomes a timelike variable everywhere in the two
``universes'', except at the matching hypersurface occupied by the \textsl{LL-brane}.
\end{abstract}


\ccode{PACS numbers: 11.25.-w, 04.70.-s,04.50.+h}

\section{Introduction}


The possibility of ``{\em non-singular} black holes'' has been pointed out
in the past few decades by numerous authors starting with the early works
\ct{non-sing-BH,bardeen}. The basic idea is to replace the standard black
hole solutions in the region containing the singularity at the center of the 
geometry ($r=0$) by regular solutions, which can be realized by introducing 
certain ``vacuum-like matter'' there \ct{markov,gonzalez}. For a systematic 
recent review and an extensive list of references, see Ref.~\refcite{ansoldi}.

Several approaches to achieve the latter
have been proposed in the literature which basically amount to:

\begin{itemize}
\item
Regular solutions of Einstein equations smoothly interpolating between
de Sitter like behaviour for small $r$ and Schwarzschild-like, or 
Reissner-Nordstr{\"o}m-like asymptotic behaviour for large $r$. This can be
achieved either by: 

(a) choosing {\em a priori} an appropriate form for the
matter energy-momentum tensor \ct{dymnikova} on the r.h.s. of Einstein equations 
without specifying a matter field lagrangian from which it could be derived, or 

(b) explicit construction of self-consistent solutions of Einstein equations, 
for gravity coupled to nonlinear electrodynamics \ct{ayon-beato,bronn}, 
\textsl{i.e.}, by deriving the pertinent energy-momentum tensor from a specific 
matter lagrangian. As shown in third Ref.~\refcite{ayon-beato}, the latter includes 
the original Bardeen regular black hole solution \ct{bardeen}.

A classification scheme for the regular black hole solutions is provided in
Ref.~\refcite{bronn-etal}.

Recently a new approach for constructing regular black holes based on non-commutative 
geometry was advocated in Refs.~\refcite{trieste-ncg}. The singularity at $r=0$ is
removed and replaced by de Sitter core due to the short-distance fluctuations
of the noncommutative coordinates.
 
Let us also note that similar regular black hole solutions with de Sitter
behavior at small distances have been previously constructed in the context
of integrable two-dimensional dilaton gravity \ct{stojkovic-etal}.
\item
Matching of interior de Sitter with exterior Schwarzschild or 
Reissner-Nordstr{\"o}m geometry across ``thin shells'', where the shell
dynamics is described in a non-Lagrangian way.
The construction in the early works \ct{gonzalez,shen-zhu}, where the matching 
occurs on the common horizon of the two geometries (\textsl{i.e.}, matching along
lightlike hypersurface), has been criticised in 
Refs.~\refcite{gron,israel-poisson-88} (see also the no-go theorem for ``false
vacuum'' black holes \ct{galtsov-lemos}). More specifically Ref.~\refcite{israel-poisson-88}
points out the occurrence of discontinuity of the pressure in Ref.~\refcite{shen-zhu} 
across the horizon separating de Sitter and Schwarzschild regions. 
On the other hand, it has been shown (Ref.~\refcite{barrabes-hogan}, p.119) that 
interior de Sitter and exterior Reissner-Nordstr{\"o}m regions can be
consistently joined along common horizon without discontinuities in the
metric and its first derivative (\textsl{i.e.}, with no $\d$-function 
singularities in the Einstein equations), provided the common horizon is the
inner horizon from the Reissner-Nordstr{\"o}m side. 

One should also mention the Frolov-Markov-Mukhanov model \ct{FMM} (see also
Ref.~\refcite{poisson-balbinot}) where de Sitter and Schwarzschild regions are matched
along spacelike spherically symmetric surface layer.
\end{itemize}

In the present paper we will pursue a different approach aimed at deriving
non-singular black holes via matching of appropriate geometries along
lightlike hypersurfaces in a self-consistent manner from first principles.
Namely, we will explore the novel possibility of employing (charged)
lightlike branes (\textsl{LL-branes}) (sometimes also called ``null branes'')
as natural self-consistent gravitational sources generating
singularity-free black hole solutions from a well-defined {\em lagrangian action 
principle} for bulk gravity-matter systems coupled to \textsl{LL-branes}.
This is achieved: 

(a) through the appearance in the self-consistent
bulk gravity-matter equations of motion of well-defined \textsl{LL-brane} 
stress-energy tensor derived from underlying reparametrization 
invariant world-volume \textsl{LL-brane} action, as well as 

(b) through dynamical generation by the \textsl{LL-brane} of space-time varying
cosmological constant. 

The systematic lagrangian action derivation of
non-singular black hole solutions through matching across \textsl{LL-brane}
is the main feature which distinguishes our present approach from the 
previously proposed approaches using ``thin shell'' matching. In particular,
as it will shown below (Eqs.\rf{pressure-cont}) the problem with pressure 
discontinuity across the horizon pointed out in Ref.~\refcite{israel-poisson-88} 
does not appear in our approach.

\textsl{LL-branes} by themselves
play an important role in general relativity as they enter the description of
various physically important cosmological and astrophysical phenomena such as: 
(i) impulsive lightlike signals arising in cataclysmic astrophysical events 
\ct{barrabes-hogan}; (ii) the ``membrane paradigm'' \ct{membrane-paradigm} of black 
hole physics; (iii) the thin-wall approach to domain walls coupled to 
gravity \ct{Israel-66,Barrabes-Israel,Dray-Hooft}.
More recently, \textsl{LL-branes} became significant also in the context of
modern non-perturbative string theory, in particular, as the so called
$H$-branes describing quantum horizons (black hole and cosmological)
\ct{kogan-01}, as Penrose limits of baryonic $D$-branes
\ct{mateos-02}, \textsl{etc} (see also Refs.~\refcite{nonperturb-string}).

In the pioneering papers \ct{Israel-66,Barrabes-Israel,Dray-Hooft} \textsl{LL-branes}
in the context of gravity and cosmology have been extensively studied from a 
phenomenological point of view, \textsl{i.e.}, by introducing them without specifying
the Lagrangian dynamics from which they may originate\foot{In a more recent paper 
\ct{barrabes-israel-05} brane actions in terms of their pertinent extrinsic geometry
have been proposed which generically describe non-lightlike branes, whereas the 
lightlike branes are treated as a limiting case.}. 
On the other hand, we have proposed in a series of recent papers 
\ct{LL-brane-main,inflation-all,our-WH} a new class of concise Lagrangian actions, 
providing a derivation from first principles of the \textsl{LL-brane} dynamics.

There are several characteristic features of \textsl{LL-branes} which drastically
distinguish them from ordinary Nambu-Goto branes: 

(i) They describe intrinsically lightlike modes, whereas Nambu-Goto branes describe
massive ones.

(ii) The tension of the \textsl{LL-brane} arises as an {\em additional
dynamical degree of freedom}, whereas Nambu-Goto brane tension is a given
{\em ad hoc} constant. 
The latter characteristic feature significantly distinguishes our \textsl{LL-brane}
models from the previously proposed {\em tensionless} $p$-branes (for a review, 
see Ref.~\refcite{lindstroem-etal}) which rather resemble a $p$-dimensional continuous
distribution of massless point-particles. 

(iii) Consistency of \textsl{LL-brane} dynamics in a spherically or axially
symmetric gravitational background of codimension one requires the presence
of an event horizon which is automatically occupied by the \textsl{LL-brane}
(``horizon straddling'' according to the terminology of 
Ref.~\refcite{Barrabes-Israel}).

(iv) When the \textsl{LL-brane} moves as a {\em test} brane in spherically or 
axially symmetric gravitational backgrounds its dynamical tension exhibits 
exponential ``inflation/deflation'' time behaviour \ct{inflation-all}
-- an effect similar to the ``mass inflation'' effect around black hole horizons
\ct{israel-poisson}. 

A simple way to realize a non-singular black hole is to consider an interior
de Sitter space-time region matched to an exterior black hole region. In
this context a particularly attractive mechanism for achieving such matching
is provided by the \textsl{LL-brane} in view of the inherent ``horizon straddling''
property of its dynamics and the dynamical (variable) nature of its tension
(properties (ii) and (iii) listed above).

The plan of the present paper is as follows. In Section 2 we consider 
self-consistent systems of bulk gravity and matter (Maxwell plus Kalb-Ramond gauge
fields) interacting with (charged) \textsl{LL-branes}. On the way we briefly 
review our construction of reparametrization invariant \textsl{LL-brane} 
world-volume actions for {\em arbitrary} world-volume dimensions. 

In Section 3 we discuss the properties of \textsl{LL-brane} equations of
motion resulting from the \textsl{LL-brane} world-volume action,
including the couplings to the bulk electromagnetic and Kalb-Ramong gauge
fields. 

In Section 4 we consider the main features of \textsl{LL-brane} dynamics in 
spherically symmetric backgrounds, namely: (a) ``horizon straddling'',
(b) producing Coulomb field in the exterior region (above the horizon),
and (c) dynamical generation of space-time varying bulk cosmological constant
(acquiring different values below and above the horizon).

In Section 5 we present the explicit construction of a solution to the
system of coupled Einstein-Maxwell-Kalb-Ramond-\textsl{LL-brane} equations 
with spherically symmetric geometry, which describes non-singular black hole.
The pertinent space-time manifold consists of two regions separated by a
codimension-one lightlike hypersurface -- the world-volume of the \textsl{LL-brane}
brane, which is simultaneously a common horizon of the interior region with de
Sitter geometry and the exterior Reissner-Nordstr{\"o}m (or 
Reissner-Nordstr{\"o}m-de-Sitter) region. Moreover, the \textsl{LL-brane}
occupying the common horizon dynamically generates the cosmological constant 
in the interior region and uniquely determines the mass and charge parameters 
of the exterior region. Here a fundamental role is being played by the dynamical 
(variable) tension of the \textsl{LL-brane}, which in this particular solution
turns out to vanish {\em on-shell}.

The resulting solution is indeed a non-singular black hole although no
violation of the weak energy condition occurs. This fact is in accordance
with the observation in Ref.~\refcite{borde} that existence of non-singular black
holes is possible provided a topology change of the spacial sections takes
place.

Section 6 is devoted to the construction of a self-consistent solution of 
Einstein-Maxwell system coupled to electrically neutral \textsl{LL-brane}, 
which describes {\em wormhole} connecting two non-singular black hole regions
(for an extensive general review of wormholes, see Ref.~\refcite{visser-book}). 
The corresponding space-time manifold consists of two identical copies of 
the exterior Reissner-Nordstr{\"o}m region above the {\em inner} 
Reissner-Nordstr{\"o}m horizon, glued together by the \textsl{LL-brane}
which automatically occupies their common horizon (the inner 
Reissner-Nordstr{\"o}m one). The corresponding mass and charge
parameters of the two Reissner-Nordstr{\"o}m ``universes'' are explicitly 
determined by the dynamical \textsl{LL-brane} tension, which in this
solution turns out to be strictly non-vanishing on-shell. The above
wormhole solution also provides an explicit example of Misner-Wheeler 
``charge without charge'' phenomenon \ct{misner-wheeler}. Here,
however, a violation of the null energy condition takes place (the 
\textsl{LL-brane} being an ``exotic matter'') as predicted by general
wormhole arguments (cf. Ref.~\refcite{visser-book}).

In the last Section 7 the wormhole solution connecting two non-singular black holes 
obtained in Section 6 is transformed into a 
cosmological bouncing solution -- a special case of Kantowski-Sachs cosmology 
\ct{KS-cosmolog}, 
by considering Reissner-Nordstr{\"o}m-de-Sitter geometry with big enough bare cosmological
constant, \textsl{i.e.}, possessing one single 
horizon. Namely,
the resulting space-time manifold consists of two identical copies of the exterior 
Reissner-Nordstr{\"o}m-de-Sitter space-time region (two identical ``universes'') glued
together along their common horizon occupied by the \textsl{LL-brane}, where
the Reissner-Nordstr{\"o}m-de-Sitter radial coordinate becomes timelike
coordinate in both ``universes'' except at the matching hypersurface (the 
\textsl{LL-brane}).

In the Appendix we describe in some detail the equivalence of \textsl{LL-brane} 
dynamics derived from the Nambu-Goto-type world-volume action 
(Eq.\rf{LL-action-NG} below) dual to the original Polyakov-type world-volume action
(Eq.\rf{LL-action} below).

\section{Lagrangian Formulation of Einstein-Maxwell-Kalb-Ramond System
Interacting With Lightlike Brane}

Self-consistent bulk Einstein-Maxwell-Kalb-Ramond system coupled to a charged 
codimension-one {\em lightlike} $p$-brane (\textsl{i.e.},
$D=(p+1)+1$) is described by the following action:
\be
S = \int\!\! d^D x\,\sqrt{-G}\,\llb \frac{R(G)}{16\pi} 
- \frac{1}{4} \cF_{\m\n}\cF^{\m\n} 
- \frac{1}{D! 2} \cF_{\m_1\ldots\m_D}\cF^{\m_1\ldots\m_D}\rrb 
+ {\wti S}_{\mathrm{LL}} \; .
\lab{E-M-KR+LL}
\ee
Here $\cF_{\m\n} = \pa_\m \cA_\n - \pa_\n \cA_\m$ and 
\be
\cF_{\m_1\ldots\m_D} = D\pa_{[\m_1} \cA_{\m_2\ldots\m_D]} =
\cF \sqrt{-G} \vareps_{\m_1\ldots\m_D}
\lab{F-KR}
\ee
are the field-strengths of the electromagnetic $\cA_\m$ and Kalb-Ramond 
$\cA_{\m_1\ldots\m_{D-1}}$ gauge potentials \ct{aurilia-townsend}.
The last term on the r.h.s. of \rf{E-M-KR+LL} indicates the reparametrization
invariant world-volume action of the \textsl{LL-brane} coupled to the bulk
gauge fields, proposed in our previous papers 
\ct{LL-brane-main,inflation-all,our-WH}:
\br
{\wti S}_{\mathrm{LL}} = S_{\mathrm{LL}}
- q \int d^{p+1}\s\,\vareps^{ab_1\ldots b_p} F_{b_1\ldots b_p} \pa_a X^\m \cA_\m
\nonu \\
- \frac{\b}{(p+1)!} \int d^{p+1}\s\,\vareps^{a_1\ldots a_{p+1}}
\pa_{a_1} X^{\m_1}\ldots\pa_{a_{p+1}} X^{\m_{p+1}} \cA_{\m_1\ldots\m_{p+1}} \; .
\lab{LL-action+EM+KR}
\er
where:
\be
S_{\mathrm{LL}} = \int d^{p+1}\s\,\P\llb -\h\g^{ab} g_{ab} + L\!\( F^2\)\rrb \; .
\lab{LL-action}
\ee
In Eqs.\rf{LL-action+EM+KR}--\rf{LL-action} the following notions and notations 
are used:

\begin{itemize}
\item
$\P$ is alternative non-Riemannian integration measure density (volume form) 
on the $p$-brane world-volume manifold:
\be
\P \equiv \frac{1}{(p+1)!} 
\vareps^{a_1\ldots a_{p+1}} H_{a_1\ldots a_{p+1}}(B) \;\; ,\;\;
H_{a_1\ldots a_{p+1}}(B) = (p+1) \pa_{[a_1} B_{a_2\ldots a_{p+1}]}
\lab{mod-measure-p}
\ee
instead of the usual $\sqrt{-\g}$. Here $\vareps^{a_1\ldots a_{p+1}}$ is the
alternating symbol ($\vareps^{0 1\ldots p} = 1$), $\g_{ab}$ ($a,b=0,1,{\ldots},p$)
indicates the intrinsic Riemannian metric on the world-volume, and
$\g = \det\Vert\g_{ab}\Vert$.
$H_{a_1\ldots a_{p+1}}(B)$ denotes the field-strength of an auxiliary
world-volume antisymmetric tensor gauge field $B_{a_1\ldots a_{p}}$ of rank $p$.
As a special case one can build $H_{a_1\ldots a_{p+1}}$ in terms of $p+1$
auxiliary world-volume scalar fields $\lcurl \vp^I \rcurl_{I=1}^{p+1}$:
\be
H_{a_1\ldots a_{p+1}} = \vareps_{I_1\ldots I_{p+1}}
\pa_{a_1} \vp^{I_1}\ldots \pa_{a_{p+1}} \vp^{I_{p+1}} \;.
\lab{mod-measure-p-scalar}
\ee
Note that $\g_{ab}$ is {\em independent} of the auxiliary world-volume fields
$B_{a_1\ldots a_{p}}$ or $\vp^I$.
The alternative non-Riemannian volume form \rf{mod-measure-p}
has been first introduced in the context of modified standard (non-lightlike) string and
$p$-brane models in Refs.~\refcite{mod-measure}.
\item
$X^\m (\s)$ are the $p$-brane embedding coordinates in the bulk
$D$-dimensional space time with bulk Riemannian metric
$G_{\m\n}(X)$ with $\m,\n = 0,1,\ldots ,D-1$; 
$(\s)\equiv \(\s^0 \equiv \t,\s^i\)$ with $i=1,\ldots ,p$;
$\pa_a \equiv \partder{}{\s^a}$.
\item
$g_{ab}$ is the induced metric on world-volume:
\be
g_{ab} \equiv \pa_a X^{\m} \pa_b X^{\n} G_{\m\n}(X) \; ,
\lab{ind-metric}
\ee
which becomes {\em singular} on-shell (manifestation of the lightlike nature, 
cf. Eq.\rf{gamma-eqs} below).
\item
$L\!\( F^2\)$ is the Lagrangian density of another
auxiliary $(p-1)$-rank antisymmetric tensor gauge field $A_{a_1\ldots a_{p-1}}$
on the world-volume with $p$-rank field-strength and its dual:
\be
F_{a_1 \ldots a_{p}} = p \pa_{[a_1} A_{a_2\ldots a_{p}]} \quad ,\quad
F^{\ast a} = \frac{1}{p!} \frac{\vareps^{a a_1\ldots a_p}}{\sqrt{-\g}}
F_{a_1 \ldots a_{p}}  \; .
\lab{p-rank}
\ee
$L\!\( F^2\)$ is {\em arbitrary} function of $F^2$ with the short-hand notation:
\be
F^2 \equiv F_{a_1 \ldots a_{p}} F_{b_1 \ldots b_{p}} 
\g^{a_1 b_1} \ldots \g^{a_p b_p} \; .
\lab{F2-id}
\ee
\end{itemize}


Rewriting the action \rf{LL-action} in the following equivalent form:
\be
S = - \int d^{p+1}\!\s \,\chi \sqrt{-\g}
\Bigl\lb \h \g^{ab} \pa_a X^{\m} \pa_b X^{\n} G_{\m\n}(X) - L\!\( F^2\)\Bigr\rb
\;\; , \; \chi \equiv \frac{\P}{\sqrt{-\g}}
\lab{LL-action-chi}
\ee
with $\P$ the same as in \rf{mod-measure-p},
we find that the composite field $\chi$ plays the role of a {\em dynamical
(variable) brane tension}. Let us note that the notion of dynamical brane 
tension has previously appeared in different contexts in Refs.~\refcite{townsend-etal}.

\textbf{Remark.}
It has been shown in Refs.~\refcite{our-WH} that the \textsl{LL-brane} equations
of motion corresponding to the Polyakov-type action \rf{LL-action} 
(or \rf{LL-action-chi}) can be equivalently obtained from the following {\em dual}
Nambu-Goto-type action:
\br
S_{\rm NG} = - \int d^{p+1}\s \, T 
\sqrt{\bgv\,\det\Vert g_{ab} - \eps \frac{1}{T^2}\pa_a u \pa_b u\Vert\,\bgv} 
\quad ,\quad \eps = \pm 1 \; .
\lab{LL-action-NG}
\er
Here $T$ is {\em dynamical} tension simply proportional to the dynamical
tension in the Polyakov-type formulation \rf{LL-action} 
~$T\sim \chi = \frac{\P}{\sqrt{-\g}}$), and $u$ denotes the dual potential 
w.r.t. $A_{a_1\ldots a_{p-1}}$:
\be
F^{\ast}_{a} (A) = \mathrm{const}\, \frac{1}{\chi} \pa_a u \; .
\lab{u-def}
\ee
Further details about the dynamical equivalence between the Polyakov-type and 
Nambu-Goto-type world-volume lagrangian formulation of \textsl{LL-branes} are 
contained in the {\em Appendix}. 
In the subsequent Sections 3--5 we will consider the original Polyakov-type 
action \rf{LL-action}.

The pertinent Einstein-Maxwell-Kalb-Ramond equations of motion derived from
the action \rf{E-M-KR+LL} read:
\br
R_{\m\n} - \h G_{\m\n} R =
8\pi \( T^{(EM)}_{\m\n} + T^{(KR)}_{\m\n} + T^{(brane)}_{\m\n}\) \; ,\phantom{aaaaa}
\lab{Einstein-eqs}  \\
\pa_\n \(\sqrt{-G}\cF^{\m\n}\) + 
q \int\!\! d^{p+1}\s\,\d^{(D)}\Bigl(x-X(\s)\Bigr)
\vareps^{ab_1\ldots b_p} F_{b_1\ldots b_p} \pa_a X^\m = 0 \; ,\phantom{aaaaa}
\lab{Maxwell-eqs}  \\
\vareps^{\n\m_1\ldots\m_{p+1}} \pa_\n \cF - \b\,\int\! d^{p+1}\s\,\d^{(D)}(x - X(\s))
\vareps^{a_1\ldots a_{p+1}} \pa_{a_1}X^{\m_1}\ldots\pa_{a_{p+1}}X^{\m_{p+1}} = 0 \; ,
\nonu \\
\phantom{aaaa}
\lab{F-KR-eqs}
\er
where in the last equation we have used relation \rf{F-KR}. 
The explicit form of the energy-momentum tensors read:
\br
T^{(EM)}_{\m\n} = \cF_{\m\k}\cF_{\n\l} G^{\k\l} - G_{\m\n}\frac{1}{4}
\cF_{\r\k}\cF_{\s\l} G^{\r\s}G^{\k\l} \; ,\phantom{aaaaa}
\lab{T-EM} \\
T^{(KR)}_{\m\n} = \frac{1}{(D-1)!}\llb \cF_{\m\l_1\ldots\l_{D-1}}
{\cF_{\n}}^{\l_1\ldots\l_{D-1}} -
\frac{1}{2D} G_{\m\n} \cF_{\l_1\ldots\l_D} \cF^{\l_1\ldots\l_D}\rrb 
= - \h \cF^2 G_{\m\n} \; ,
\nonu \\
\phantom{aaaa}
\lab{T-KR} \\
T^{(brane)}_{\m\n} = - G_{\m\k}G_{\n\l}
\int\!\! d^{p+1}\s\,\frac{\d^{(D)}\Bigl(x-X(\s)\Bigr)}{\sqrt{-G}}\,
\chi\,\sqrt{-\g} \g^{ab}\pa_a X^\k \pa_b X^\l \; ,\phantom{aaaaa}
\lab{T-brane}
\er
where the brane stress-energy tensor is straightforwardly derived
from the world-volume action \rf{LL-action} (or, equivalently, \rf{LL-action-chi};
recall $\chi\equiv\frac{\P}{\sqrt{-\g}}$ is the variable brane tension).

Eqs.\rf{Maxwell-eqs}--\rf{F-KR-eqs} show that:

(i) the \textsl{LL-brane} is charged source for the bulk electromagnetism;

(ii) the \textsl{LL-brane} uniquely determines the
value of $\cF^2$ in Eq.\rf{T-KR} through its coupling to the bulk
Kalb-Ramond gauge field (Eq.\rf{F-KR-eqs}) which implies {\em dynamical generation} of
bulk cosmological constant $\L=4\pi \cF^2$.

The equations of motion of the \textsl{LL-brane} are discussed in some
detail in the next Section.

\section{Lightlike Brane Dynamics}

The equations of motion derived from the brane action \rf{LL-action} w.r.t. 
$B_{a_1\ldots a_{p}}$ are:
\be
\pa_a \Bigl\lb \h \g^{cd} g_{cd} - L(F^2)\Bigr\rb = 0 \quad \longrightarrow \quad
\h \g^{cd} g_{cd} - L(F^2) = M  \; ,
\lab{phi-eqs}
\ee
where $M$ is an arbitrary integration constant. The equations of motion w.r.t.
$\g^{ab}$ read:
\be
\h g_{ab} - F^2 L^{\pr}(F^2) \llb\g_{ab} 
- \frac{F^{*}_a F^{*}_b}{F^{*\, 2}}\rrb = 0  \; ,
\lab{gamma-eqs}
\ee
where $F^{*\, a}$ is the dual world-volume field strength \rf{p-rank}. 

\textbf{Remark 1.} Before proceeding, let us mention that both the auxiliary 
world-volume $p$-form gauge field $B_{a_1\ldots a_{p}}$ entering the 
non-Riemannian integration measure density \rf{mod-measure-p}, as well as the
intrinsic world-volume metric $\g_{ab}$ are
{\em non-dynamical} degrees of freedom in the action \rf{LL-action},
or equivalently, in \rf{LL-action-chi}. Indeed, there are no (time-)derivatives
w.r.t. $\g_{ab}$, whereas the action \rf{LL-action} (or \rf{LL-action-chi}) is
{\em linear} w.r.t. the velocities $\pa_0 B_{a_1\ldots a_{p}}$. Thus,
\rf{LL-action} is a constrained dynamical system, \textsl{i.e.}, a system with 
gauge symmetries including the gauge symmetry under world-volume
reparametrizations, and 
both Eqs.\rf{phi-eqs}--\rf{gamma-eqs} are in fact
{\em non-dynamical constraint} equations (no second-order time derivatives
present). Their meaning as constraint equations is best understood within the 
framework of the canonical Hamiltonian formalism for the action \rf{LL-action}. 
Using the latter formalism one can show that also the auxiliary world-volume 
$(p-1)$-form gauge field $A_{a_1\ldots a_{p-1}}$ \rf{p-rank} is non-dynamical.
The canonical Hamiltonian formalism for the action \rf{LL-action} can be developed
in strict analogy with the Hamiltonian formalism for a simpler class of
modified {\em non-lightlike} $p$-brane models based on the alternative 
non-Riemannian integration measure density \rf{mod-measure-p}, which was 
previously proposed in Ref.~\refcite{m-string} (for details, we refer to Sections 2 and 3
of Ref.~\refcite{m-string}). In particular, Eqs.\rf{gamma-eqs} can be viewed as $p$-brane 
analogues of the string Virasoro constraints.

There are two important consequences of Eqs.\rf{phi-eqs}--\rf{gamma-eqs}.
Taking the trace in \rf{gamma-eqs} and comparing with \rf{phi-eqs} 
implies the following crucial relation for the Lagrangian function $L\( F^2\)$: 
\be
L\!\( F^2\) - p F^2 L^\pr\!\( F^2\) + M = 0 \; ,
\lab{L-eq}
\ee
which determines $F^2$ \rf{F2-id} on-shell as certain function of the integration
constant $M$ \rf{phi-eqs}, \textsl{i.e.}
\be
F^2 = F^2 (M) = \mathrm{const} \; .
\lab{F2-const}
\ee

The second and most profound consequence of Eqs.\rf{gamma-eqs} is that the induced 
metric \rf{ind-metric} on the world-volume of the $p$-brane model \rf{LL-action} 
is {\em singular} on-shell (as opposed to the induced metric in the case of 
ordinary Nambu-Goto branes):
\be
g_{ab}F^{*\, b}=0 \; ,
\lab{on-shell-singular}
\ee
\textsl{i.e.}, the tangent vector to the world-volume $F^{*\, a}\pa_a X^\m$
is {\em lightlike} w.r.t. metric of the embedding space-time.
Thus, we arrive at the following important conclusion: every point on the 
surface of the $p$-brane \rf{LL-action} moves with the speed of light
in a time-evolution along the vector-field $F^{\ast a}$ which justifies the
name {\em LL-brane} (lightlike brane) model for \rf{LL-action}.

\textbf{Remark 2.} Let us stress the importance of introducing the alternative 
non-Riemannian integration measure density in the form \rf{mod-measure-p}.
If we would have started with world-volume \textsl{LL-brane} action in the
form \rf{LL-action-chi} where the tension $\chi$ would be an {\em elementary}
scalar field (instead of being a composite one -- a ratio of two scalar
densities as in the second relation in \rf{LL-action-chi}), then variation
w.r.t. $\chi$ would produce second Eq.\rf{phi-eqs} with $M$ identically zero. This
in turn by virtue of the constraint \rf{L-eq} (with $M=0$) would require the
Lagrangian $L(F^2)$ to assume the special fractional power function form
$L(F^2) = \( F^2\)^{1/p}$. In this special case the action \rf{LL-action-chi} with
elementary field $\chi$ becomes in addition manifestly invariant under 
{\em Weyl (conformal) symmetry}: 
$\g_{ab}\!\longrightarrow\! \g^{\pr}_{ab}\! =\! \rho\,\g_{ab}$,
$\chi \! \longrightarrow\! \chi^{\pr}\! =\! \rho^{\frac{1-p}{2}}\chi$. This special 
case of Weyl-conformally invariant \textsl{LL-branes} has been discussed in our 
older papers (first two Refs.~\refcite{LL-brane-main}).

Let us point out that supplementing the \textsl{LL-brane} 
action \rf{LL-action} with natural couplings to bulk Maxwell and 
Kalb-Ramond gauge fields, as explicitly given by the world-volume action
in Eq.\rf{LL-action+EM+KR}, does not affect Eqs.\rf{phi-eqs} and 
\rf{gamma-eqs}, so that the conclusions about on-shell constancy of $F^2$ 
\rf{F2-const} and the lightlike nature \rf{on-shell-singular} of the $p$-branes 
under consideration remain unchanged. In what follows we will consider the
extended \textsl{LL-brane} world-volume action \rf{LL-action+EM+KR}.

It remains to write down the equations of motion w.r.t. auxiliary
world-volume gauge field $A_{a_1 \ldots a_{p-1}}$ and $X^\m$ produced by the
\textsl{LL-brane} action \rf{LL-action+EM+KR}:
\br
\pa_{[a}\( F^{\ast}_{b]}\, \chi L^\pr (F^2)\) 
+ \frac{q}{4}\pa_a X^\m \pa_b X^\n \cF_{\m\n} = 0  \; ;\phantom{aaaaa}
\lab{A-eqs} \\
\pa_a \(\chi \sqrt{-\g} \g^{ab} \pa_b X^\m\) + 
\chi \sqrt{-\g} \g^{ab} \pa_a X^\n \pa_b X^\l \G^\m_{\n\l}
-q \vareps^{ab_1\ldots b_p} F_{b_1\ldots b_p} \pa_a X^\n \cF_{\l\n}G^{\l\m}
\nonu \\
- \frac{\b}{(p+1)!} \vareps^{a_1\ldots a_{p+1}} \pa_{a_1} X^{\m_1} \ldots
\pa_{a_{p+1}} X^{\m_{p+1}} \cF_{\l\m_1\dots\m_{p+1}} G^{\l\m} = 0 \; .\phantom{aaaaa}
\lab{X-eqs}
\er
Here $\chi$ is the dynamical brane tension as in \rf{LL-action-chi}, 
$\cF_{\l\m_1\dots\m_{p+1}}$ is the Kalb-Ramond field-strength \rf{F-KR},
\be
\G^\m_{\n\l}=\h G^{\m\k}\(\pa_\n G_{\k\l}+\pa_\l G_{\k\n}-\pa_\k G_{\n\l}\)
\lab{affine-conn}
\ee
is the Christoffel connection for the external metric,
and $L^\pr(F^2)$ denotes derivative of $L(F^2)$ w.r.t. the argument $F^2$.

World-volume reparametrization invariance 
allows to introduce the standard synchronous gauge-fixing conditions:
\be
\g^{0i} = 0 \;\; (i=1,\ldots,p) \; ,\; \g^{00} = -1 \; .
\lab{gauge-fix}
\ee
Also, we will use a natural ansatz for the ``electric'' part of the 
auxiliary world-volume gauge field-strength \rf{p-rank}:
\be
F^{\ast i}= 0 \;\; (i=1,{\ldots},p) \quad ,\quad \mathrm{i.e.} \;\;
F_{0 i_1 \ldots i_{p-1}} = 0 \; ,
\lab{F-ansatz}
\ee
meaning that we choose the lightlike direction in Eq.\rf{on-shell-singular} 
to coincide with the brane
proper-time direction on the world-volume ($F^{*\, a}\pa_a \sim \pa_\t$).
The Bianchi identity ($\nabla_a F^{\ast\, a}=0$) together with 
\rf{gauge-fix}--\rf{F-ansatz} and the definition for the dual field-strength
in \rf{p-rank} imply:
\be
\pa_0 \g^{(p)} = 0 \quad \mathrm{where}\;\; \g^{(p)} \equiv \det\Vert\g_{ij}\Vert \; .
\lab{gamma-p-0}
\ee
Then the \textsl{LL-brane} equations of motion acquire the form 
(recall definition of the induced metric $g_{ab}$ \rf{ind-metric}):
\be
g_{00}\equiv \Xdot^\m\!\! G_{\m\n}\!\! \Xdot^\n = 0 \quad ,\quad g_{0i} = 0 \quad ,\quad
g_{ij} - 2a_0\, \g_{ij} = 0
\lab{gamma-eqs-0}
\ee
(the latter are analogs of Virasoro constraints in standard string theory), where 
$a_0$ is a $M$-dependent constant:
\be
a_0 \equiv F^2 L^\pr (F^2)\bv_{F^2 = F^2(M)}  \; ;
\lab{a0-const}
\ee
\be
\pa_i \chi + \frac{q}{a_1}\pa_0 X^\m \pa_i X^\n \cF_{\m\n} = 0 \quad ,\quad
\pa_i X^\m \pa_j X^\n \cF_{\m\n} = 0 \; ,
\lab{A-eqs-0}
\ee
with
\be
a_1 \equiv 2\sqrt{F^2/p!} L^\pr (F^2)\bv_{F^2 = F^2(M)} = \mathrm{const}
\lab{a1-const} \; ;
\ee
\br
-\sqrt{\g^{(p)}} \pa_0 \(\chi \pa_0 X^\m\) +
\pa_i\(\chi\sqrt{\g^{(p)}} \g^{ij} \pa_j X^\m\) \phantom{a}
\nonu \\
+ \chi\sqrt{\g^{(p)}} \(-\pa_0 X^\n \pa_0 X^\l + \g^{kl} \pa_k X^\n \pa_l X^\l\)
\G^\m_{\n\l} - q\,\sqrt{p!\, F^2} \sqrt{\g^{(p)}} \pa_0 X^\n \cF_{\l\n}G^{\l\m}
\phantom{a}
\nonu \\
- \frac{\b}{(p+1)!}\cF \vareps^{a_1\ldots a_{p+1}} \pa_{a_1} X^{\m_1} \ldots
\pa_{a_{p+1}} X^{\m_{p+1}} \vareps_{\l \m_1 \dots \m_{p+1}} \sqrt{-G} G^{\l\m} = 0 \; .
\phantom{a}
\lab{X-eqs-0}
\er
Let us recall that $F^2 = \mathrm{const}$ (Eq.\rf{F2-const}).

\section{Lightlike Brane in Spherically Symmetric Backgrounds}

In what follows we will be interested in static spherically symmetric solutions 
of Einstein-Maxwell-Kalb-Ramond equations \rf{Einstein-eqs}--\rf{F-KR-eqs}.
The generic form of spherically symmetric metric in Eddington-Finkelstein 
coordinates \ct{EFM} reads:
\be
ds^2 = - A(r) dv^2 + 2 dv\,dr + C(r) h_{ij}(\th) d\th^i d\th^j \; ,
\lab{EF-metric}
\ee
where $h_{ij}$ indicates the standard metric on $S^p$.
We will consider the simplest ansatz for the \textsl{LL-brane} embedding
coordinates:
\be
X^0\equiv v = \t \quad, \quad X^1\equiv r = r(\t) \quad, \quad 
X^i\equiv \th^i = \s^i \;\; (i=1,\ldots ,p)
\lab{X-embed}
\ee
Now, the \textsl{LL-brane} equations \rf{gamma-eqs-0} together with \rf{gamma-p-0}
yield:
\be
-A(r) + 2\rdot = 0 \quad , \quad \pa_\t C = \rdot\,\pa_r C\bv_{r=r(\t)} = 0 \; ,
\lab{r-const}
\ee
implying:
\be
\rdot = 0 \; \to \; r = r_0 = \mathrm{const} \quad ,\quad A(r_0) = 0 \; .
\lab{horizon-standard}
\ee
Eq.\rf{horizon-standard} tells us that consistency of \textsl{LL-brane} dynamics in 
a spherically symmetric gravitational background of codimension one requires the 
latter to possess a horizon (at some $r = r_0$), which is automatically occupied 
by the \textsl{LL-brane} (``horizon straddling'' according to the
terminology of Ref.~\refcite{Barrabes-Israel}). Similar property -- 
``horizon straddling'', has been found also for \textsl{LL-branes} moving in
rotating axially symmetric (Kerr or Kerr-Newman) and rotating cylindrically
symmetric black hole backgrounds (first and third Ref.~\refcite{our-WH}).

Next, the Maxwell coupling of the \textsl{LL-brane} produces via 
Eq.\rf{Maxwell-eqs} static Coulomb field in the outer region beyond the
horizon. Namely, inserting in Eq.\rf{Maxwell-eqs}
the embedding ansatz \rf{X-embed} together with \rf{horizon-standard} 
and accounting for \rf{gauge-fix}--\rf{gamma-eqs-0} we obtain;
\be
\pa_r\( C^{p/2}(r)\cF_{vr} (r)\) - 
q\,\frac{\sqrt{p! F^2}}{(2a_0)^{p/2}} C^{p/2}(r_0) \d (r-r_0) = 0 \; ,
\lab{Maxwell-0}
\ee
which yields for the Maxwell field-strength:
\be
\cF_{vr} (r) = \(\frac{C(r_0)}{C(r)}\)^{p/2} 
\frac{q\sqrt{p! F^2}}{(2a_0)^{p/2}}\th(r-r_0) \; .
\lab{F-Maxwell}
\ee
Taking into account the Coulomb form of the bulk Maxwell field-strength
\rf{F-Maxwell} together with \rf{X-embed} and \rf{horizon-standard}, we
obtain that \textsl{LL-brane} Eqs.\rf{A-eqs-0} reduce to simply $\pa_i \chi = 0$.

Using again the embedding ansatz \rf{X-embed} together with \rf{horizon-standard}
as well as \rf{gauge-fix}--\rf{gamma-eqs-0}, the Kalb-Ramond equations of motion
\rf{F-KR-eqs} reduce to:
\br
\pa_r \cF + \b \d (r-r_0) = 0 \quad \to \quad 
\cF = \cF_{(+)} \th (r-r_0) + \cF_{(-)} \th (r_0 -r)
\lab{F-KR-0}\\
\cF_{(\pm)} = \mathrm{const} \quad ,\quad \cF_{(-)} - \cF_{(+)} = \b
\lab{F-jump}
\er
Therefore, a space-time varying non-negative cosmological constant is dynamically 
generated in both exterior and interior regions w.r.t. the horizon at $r=r_0$
(cf. Eq.\rf{T-KR}):
\be
\L_{(\pm)} = 4\pi \cF^2_{(\pm)} \; .
\lab{cosmolog-const}
\ee

Finally, it remains to consider the second order (w.r.t. proper time derivative) 
$X^\m$ equations of motion \rf{X-eqs-0}. Upon inserting the embedding ansatz
\rf{X-embed} together with \rf{horizon-standard} and taking into account
\rf{gamma-eqs-0}, \rf{F-Maxwell} and \rf{F-jump}, we find that the only non-trivial
equations is for $\m=v$. Before proceeding let us note that the ``force''
terms in the $X^\m$ equations of motion \rf{X-eqs-0} (the geodesic ones
containing the Christoffel connection coefficients as well as those coming
from the \textsl{LL-brane} coupling to the bulk Maxwell and Kalb-Ramond
gauge fields) contain discontinuities across the horizon occupied by the
\textsl{LL-brane}. The discontinuity problem is resolved following 
the approach in Ref.~\refcite{Israel-66} (see also the regularization
approach in Ref.~\refcite{BGG}, Appendix A) by taking mean values of the ``force''
terms across the discontinuity at $r=r_0$. Thus, we obtain from Eq.\rf{X-eqs-0}
with $\m=v$:
\br
\pa_\t \chi + \chi \llb \frac{1}{4}\(\pa_r A_{(+)} + \pa_r A_{(-)}\) 
+ p a_0 \pa_r \ln C \rrb_{r=r_0}
\nonu \\
+ \h \Bigl\lb - q^2\frac{p! F^2}{(2a_0)^{p/2}}
+ \b (2a_0)^{p/2} \(\cF_{(-)} + \cF_{(+)}\)\Bigr\rb = 0
\lab{X-0-eq} \; .
\er


\section{Non-Singular Black Hole Solution}

Let us go back to the Einstein equations of motion \rf{Einstein-eqs}, where
the \textsl{LL-brane} energy-momentum tensor \rf{T-brane}, upon inserting the 
expressions for $X^\m (\s)$ from \rf{X-embed} and \rf{horizon-standard}, 
and taking into account \rf{gauge-fix}, \rf{gamma-p-0} and \rf{gamma-eqs-0}, 
acquires the form:
\be
T_{(brane)}^{\m\n} = S^{\m\n}\,\d (r-r_0)
\lab{T-S-0}
\ee
with surface energy-momentum tensor:
\be
S^{\m\n} \equiv \frac{\chi}{(2a_0)^{p/2}}\,
\llb \pa_\t X^\m \pa_\t X^\n - 2a_0 G^{ij} \pa_i X^\m \pa_j X^\n 
\rrb_{v=\t,\,r=r_0,\,\th^i =\s^i} \; .
\lab{T-S-brane}
\ee
Here again $a_0$ is the integration constant parameter appearing in the 
\textsl{LL-brane} dynamics \rf{a0-const} and $G_{ij} = C(r) h_{ij}(\th)$.
For the non-zero components of $S_{\m\n}$ (with lower indices) and its trace we find:
\be
S_{rr} = \frac{\chi}{(2a_0)^{p/2}} \quad ,\quad 
S_{ij} = - \frac{\chi}{(2a_0)^{p/2-1}} G_{ij} \quad ,\quad
S^\l_\l = - \frac{p\chi}{(2a_0)^{p/2-1}} 
\lab{S-comp}
\ee
The solution of the other bulk space-time equations of motion (the Maxwell
\rf{Maxwell-eqs} and Kalb-Ramond \rf{F-KR-eqs}) with spherically symmetric
geometry have already been given in the previous Section, see
Eqs.\rf{Maxwell-0}--\rf{F-jump}.

For the sake of simplicity we will consider in what follows the case of 
$D=4$-dimensional bulk space-time and, correspondingly, $p=2$ for the 
\textsl{LL-brane}. The generalization to arbitrary $D$ is straightforward.
For further simplification of the numerical constant factors we will choose
the following specific form for the Lagrangian of the auxiliary non-dynamical
world-volume gauge field (cf. Eqs.\rf{p-rank}--\rf{F2-id}): 
\be
L(F^2)=\frac{1}{4}F^2 \quad \to \quad  a_0 = M \; ,
\lab{L-eq-0}
\ee
where $a_0$ is the constant defined in \rf{a0-const} and
$M$ denotes the original integration constant in Eqs.\rf{phi-eqs} and \rf{L-eq}.

We will show that there exists a spherically symmetric solution of the
Einstein equations of motion \rf{Einstein-eqs} with \textsl{LL-brane}
energy-momentum tensor on the r.h.s. given by \rf{T-S-brane}--\rf{S-comp} --
systematically derived from the reparametrization invariant \textsl{LL-brane}
world-volume action \rf{LL-action}, which describes a {\em non-singular}
black hole. Namely, this solutions describes a space-time consisting of two
spherically symmetric regions -- an interior de-Sitter region (for $r< r_0$)
and an exterior Reissner-Nordstr{\"o}m (or Reissner-Nordstr{\"o}m-de-Sitter) region
(for $r > r_0$) matched along common horizon at $r=r_0$, where $r_0$ is the
{\em inner} horizon from the Reissner-Nordstr{\"o}m side and which is automatically 
occupied by the \textsl{LL-brane} (``horizon straddling'') as shown in 
Eqs.\rf{r-const}--\rf{horizon-standard} above. Moreover:

(a) The surface charge density $q$ of the \textsl{LL-brane} (cf. \rf{Maxwell-eqs}
and \rf{Maxwell-0})
explicitly determines the non-zero Coulomb field-strength in the exterior region
$r > r_0$ (Eq.\rf{F-Maxwell}) so that the Reissner-Nordstr{\"o}m charge parameter 
is given explicitly by (for $D=4$ space-time dimensions):
\be
Q^2 = \frac{8\pi}{a_0} q^2\, r_0^4 \; .
\lab{RN-charge}
\ee

(b) As shown in Eqs.\rf{F-KR-0}--\rf{cosmolog-const}, The \textsl{LL-brane} 
through its coupling to the bulk Kalb-Ramond field (cf. \rf{LL-action+EM+KR}
and \rf{F-KR-eqs}) dynamically generates space-time varying non-negative
cosmological constant with a jump across the horizon ($r=r_0$). In particular, 
we will consider the case of vanishing cosmological constant in the exterior region
($r > r_0$) (pure Reissner-Nordstr{\"o}m geometry) which leaves a dynamically 
generated de Sitter geometry below the horizon ($r < r_0$),
\textsl{i.e.}, $\L_{(-)} = 4\pi \b^2$.

In Eddington-Finkelstein coordinates the non-singular black hole is given by:
\br
ds^2 = - A(r) dv^2 + 2 dv\,dr + r^2 h_{ij}(\th) d\th^i d\th^j \; .
\lab{EF-metric-1} \\
A(r) \equiv A_{(-)}(r) = 1 - \frac{4\pi}{3}\cF^2_{(-)} r^2 \quad , \;\; 
\mathrm{for}\; r < r_0 \; ;
\lab{de-Sitter}\\
A(r) \equiv A_{(+)}(r) = 1 - \frac{2m}{r} + \frac{Q^2}{r^2} - 
\frac{4\pi}{3}\cF^2_{(+)} r^2 \quad , \;\; \mathrm{for}\;  r > r_0 \; ,
\lab{RN-dS}
\er
where $\cF_{(\pm)}$ and $Q^2$ is given by \rf{F-jump} and \rf{RN-charge}, 
respectively, and where $r_0$ is the common horizon:
\be
A_{(-)}(r_0) = 0 \quad ,\quad A_{(+)}(r_0) = 0 \quad ,\quad 
r_0 = m - \sqrt{m^2 - Q^2}\;\; (\,\mathrm{for}\; \cF_{(+)}=0\,) \;.
\lab{common-horizon}
\ee

Since the above metrics obviously solve Einstein equations \rf{Einstein-eqs}
in their respective interior ($r < r_0$) and exterior ($r > r_0$) regions, 
it only remains to study their matching along the common horizon ($r = r_0$).

The systematic formalism for matching different bulk space-time geometries on
codimension-one timelike hypersurfaces (``thin shells'') was developed 
originally in Ref.~\refcite{Israel-66} and later generalized in 
Ref.~\refcite{Barrabes-Israel} 
to the case of lightlike hypersurfaces (``null thin shells'') (for a
systematic introduction, see the textbook \ct{poisson-kit}).
In the present case, due to the simple geometry (spherical symmetry and
matching on common horizon) one can straightforwardly isolate the terms from
the Ricci tensor on the l.h.s. of Einstein equations \rf{Einstein-eqs} which
may yield delta-function contributions ($\sim \d (r - r_0)$) to be matched
with the components of the \textsl{LL-brane} surface stress-energy tensor
\rf{T-S-0}--\rf{T-S-brane}.

Since the metric \rf{EF-metric-1}--\rf{RN-dS} is continuous at $r=r_0$
(due to Eq.\rf{common-horizon}), but its first derivative w.r.t. $r$ (the normal
coordinate w.r.t. horizon) might exhibit discontinuity across $r=r_0$,
the terms contributing to $\delta$-function singularities in $R_{\m\n}$
are those containing second derivatives w.r.t. $r$. Separating explicitly
the latter we can rewrite Eqs.\rf{Einstein-eqs} in the following form
(here we take $D=4$ and $p=2$):
\br
R_{\m\n} \equiv \pa_r \G^r_{\m\n} - \pa_\m \pa_\n \ln \sqrt{-G}
+ \mathrm{non-singular ~terms}
\nonu \\
= 8\pi \( S_{\m\n} - \h G_{\m\n} S^{\l}_{\l}\) \d (r-r_0) 
+ \mathrm{non-singular ~terms} \; .
\lab{Einstein-eqs-sing}
\er
For $(\m,\n) = (r,r)$ the r.h.s. of Eq.\rf{Einstein-eqs-sing} has non-zero
$\delta$-function contribution due to the non-zero component 
$S_{rr}= \frac{\chi}{2a_0}$ (cf. Eqs.\rf{S-comp}) while on the l.h.s. 
$\G^r_{rr}=0$ for the metric \rf{EF-metric-1} and $\sqrt{-G}=r^2$ is a smooth
function across $r=r_0$, \textsl{i.e.}, there is no $\d (r-r_0)$
contribution on the l.h.s. Therefore, the matching dictates that the
dynamical \textsl{LL-brane} tension $\chi$ must vanish on-shell in the
solution under consideration.

Then, Eq.\rf{Einstein-eqs-sing} for $(\m,\n) = (v,r)$ yields:
\be
\pa_r A_{(+)} \bv_{r=r_0} - \pa_r A_{(-)} \bv_{r=r_0} = 0 \; .
\lab{deriv-metric-match}
\ee
Let us note that in spite of the continuity of the metric \rf{EF-metric-1}
and its first derivative w.r.t. $r$ across the horizon $(r=r_0)$ 
(Eqs.\rf{common-horizon} and \rf{deriv-metric-match}), the matching is 
{\em not} completely smooth since the higher derivatives of \rf{EF-metric-1} are 
discontinuous there.

Finally, taking into account that $\chi=0$ on-shell, Eq.\rf{X-0-eq} yields a
second relation between the dynamically generated cosmological constants
below and above the horizon (here $p=2$):
\be 
\cF_{(-)} + \cF_{(+)} = \frac{2q^2}{a_0 \b} \; ,
\lab{cF-sum}
\ee
with together with \rf{F-jump} completely determines $\cF_{(\pm)}$ as
functions of the \textsl{LL-brane} charge and Kalb-Ramond coupling parameters:
\be
\cF_{(\pm)} = \frac{q^2}{a_0 \b} \mp \frac{\b}{2} \; .
\lab{cF-eqs}
\ee

Solving the matching equations \rf{common-horizon}, \rf{deriv-metric-match}
we get:
\be
r_0 = \frac{1}{\sqrt{K_{(-)}}} \;\; ,\;\; 
m = \frac{2}{\sqrt{K_{(-)}}} \( 1 - \frac{K_{(+)}}{K_{(-)}}\) \;\; ,\;\;
Q^2 = \frac{3}{K_{(-)}} \( 1 - \frac{K_{(+)}}{K_{(-)}}\) \; ,
\lab{metric-matching}
\ee
where (cf. Eqs.\rf{cF-eqs}):
\be
K_{(\pm)} \equiv \frac{4\pi}{3} \cF^2_{(\pm)} =
\frac{4\pi}{3} \(\frac{q^2}{a_0 \b} \mp \frac{\b}{2}\)^2 \; ,
\lab{K-eqs}
\ee
and the expression for $Q^2$ in \rf{metric-matching} is identical to that 
in Eq.\rf{RN-charge}.

In particular, choosing the \textsl{LL-brane} integration constant
$a_0 = 2 \frac{q^2}{\b^2}$ we have from \rf{cF-eqs}:
\be
\cF_{-} = \b \quad ,\quad \cF_{(+)} = 0 \; ,
\lab{cF-eqs-0}
\ee
\textsl{i.e.}, vanishing dynamical cosmological constant above horizon.
In this special case of interior de Sitter region matched to pure
Reissner-Nordstr{\"o}m exterior region Eqs.\rf{metric-matching} simplify to:
\be
r_0 = \frac{1}{\sqrt{K}} \quad ,\quad
m = \frac{2}{\sqrt{K}} \quad ,\quad
Q^2 = \frac{3}{K} \quad ,\quad \mathrm{with}\;\; K = \frac{4\pi}{3} \b^2 \; .
\lab{metric-matching-1}
\ee

It is now straightforward to check the {\em absence} of pressure discontinuity across
the matching horizon. Indeed, using the relations \rf{F-Maxwell} and 
\rf{cosmolog-const}, \rf{cF-eqs} for the \textsl{LL-brane}-generated exterior 
Coulomb field-strength and space-time-varying cosmological constant and
inserting them into the expressions for the corresponding energy-momentum
tensors \rf{T-EM}--\rf{T-KR} we obtain for the mixed diagonal $({}^r_r)$ components 
of the latter at the common horizon:
\be
(T^{\mathrm{interior}}_{\mathrm{dS}})^r_r = 
(T^{\mathrm{exterior}}_{\mathrm{RN+dS}})^r_r =
- \h \(\frac{q^2}{a_0 \b} + \frac{\b}{2}\)^2 \; .
\lab{pressure-cont}
\ee
Similar conclusion about absence of discontinuity in the energy density
across the matching horizon is true as well, since in the present case
$T^0_0 = T^r_r$ in both regions.

\section{Wormhole Connecting Two Non-Singular Black Holes}

Let us now consider a self-consistent bulk Einstein-Maxwell system ({\em without} 
cosmological constant) free of electrically charged matter, coupled to a 
electrically neutral \textsl{LL-brane}:
\be
S = \int\!\! d^D x\,\sqrt{-G}\,\llb \frac{R(G)}{16\pi}
- \frac{1}{4} \cF_{\m\n}\cF^{\m\n}\rrb + S_{\mathrm{LL}} \; ,
\lab{E-M-LL}
\ee
where $S_{\mathrm{LL}}$ is the same as in \rf{LL-action} (Polyakov-type 
\textsl{LL-brane} action) or in \rf{LL-action-NG-A} (Nambu-Goto-type \textsl{LL-brane}
action). The pertinent Einstein-Maxwell equations of motion read:
\be
R_{\m\n} - \h G_{\m\n} R  =
8\pi \( T^{(EM)}_{\m\n} + T^{(brane)}_{\m\n}\) \quad, \quad
\pa_\n \(\sqrt{-G}G^{\m\k}G^{\n\l} \cF_{\k\l}\) = 0 \; ,
\lab{Einstein-Maxwell-eqs}
\ee
where $T^{(EM)}_{\m\n}$ is given by \rf{T-EM} and $T^{(brane)}_{\m\n}$ is
the same as in \rf{T-brane} or, equivalently, \rf{T-brane-A}.

Using the general formalism for constructing self-consistent spherically
symmetric or rotating cylindrical {\em wormhole} solutions via \textsl{LL-branes} 
presented in third Ref.~\refcite{our-WH}, we will now construct traversable
(w.r.t. the proper time of travelling observer, see below) wormhole solution to 
the Einstein equations \rf{Einstein-Maxwell-eqs} which: 

(a) will connect two identical non-singular black hole space-time regions -- 
two copies of the exterior Reissner-Nordstr{\"o}m region above the {\em inner} 
Reissner-Nordstr{\"o}m horizon; 

(b) will combine the 
features of the Einstein-Rosen ``bridge'' in its original formulation
\ct{einstein-rosen} (wormhole throat at horizon \ct{einstein-rosen-1})
%
and the feature ``charge without charge'' of Misner-Wheeler wormholes 
\ct{misner-wheeler}.

In doing this we will follow the standard procedure described in
\ct{visser-book}, but with the significant difference that in our case
we will solve Einstein equations following from a well-defined lagrangian action
(\rf{E-M-LL} with \rf{LL-action} or \rf{LL-action-NG-A}) describing 
self-consistent bulk gravity-matter system coupled to a \textsl{LL-brane}.
In other words, the \textsl{LL-brane} will serve as a gravitational source
of the wormhole by locating itself on its throat as a result
of its consistent world-volume dynamics (Eq.\rf{horizon-standard} above). 

Let us introduce the following modification of the standard 
Reissner-Nordstr{\"o}m metric in Eddington-Finkelstein coordinates::
\br
ds^2 = - {\wti A}(\eta) dv^2 + 2 dv d\eta +
{\wti r}^2 (\eta) h_{ij}(\vec{\th}) d\th^i d\th^j \; ,
\lab{RN-W-1} \\
{\wti A}(\eta) \equiv A\bigl({\wti r}(\eta)\bigr)=
1 - \frac{2m}{{\wti r}(\eta)} + \frac{Q^2}{{\wti r}^2(\eta)}
\quad , \quad {\wti r}(\eta) = r_{(-)} + |\eta| \;,
\lab{RN-W-2}
\er
where:
\be
r_{(-)} \equiv m - \sqrt{m^2 - Q^2} \quad ,\quad -\infty < \eta < \infty \; . 
\lab{r-eta}
\ee
From now on the bulk space-time indices $\m,\n$ will refer to $(v,\eta,\th^i)$
instead of $(v,r,\th^i)$.
The new metric \rf{RN-W-1}--\rf{RN-W-2} represents two identical copies of the
exterior Reissner-Nordstr{\"o}m space-time region $(r > r_{(-)})$, which are 
sewed together along the internal Reissner-Nordstr{\"o}m horizon $(r=r_{(-)})$
\rf{r-eta}.
We will show that the new metric \rf{RN-W-1}--\rf{RN-W-2} is a solution of the
Einstein equations \rf{Einstein-Maxwell-eqs} thanks to the presence of  
$T^{(brane)}_{\m\n}$ on the r.h.s.. Here the newly introduced coordinate $\eta$
will play the role of a radial-like coordinate normal w.r.t. the \textsl{LL-brane}
located on the throat, which interpolates between the two Reissner-Nordstr{\"o}m
``universes'' for $\eta > 0$ and  $\eta < 0$ (the two copies 
transform into each other under the ``parity'' transformation $\eta \to - \eta$).
Each of these Reissner-Nordstr{\"o}m ``universes'' represents a non-singular
black hole space-time region since they still contain horizons at
$\eta = 2 \sqrt{m^2 - Q^2}$ and $\eta = - 2 \sqrt{m^2 - Q^2}$, respectively,
which correspond to the outer horizon $r_{(+)} \equiv m + \sqrt{m^2 - Q^2}$
of the standard Reissner-Nordstr{\"o}m metric.

As in \rf{X-embed} we will use the simplest embedding for the \textsl{LL-brane} 
coordinates $\bigl( X^\m\bigr) \equiv (v,\eta,\th^i) = (\t, \eta (\t),\s^i)$.
In complete analogy with \rf{r-const}--\rf{horizon-standard} we find that
the \textsl{LL-brane} equations of motion following from the underlying 
world-volume action \rf{LL-action} or, equivalently, \rf{LL-action-NG} 
yield $\eta (\t)=0$, \textsl{i.e.}, the \textsl{LL-brane} automatically
locates itself on the junction $(\eta = 0)$ between the two
Reissner-Nordstr{\"o}m ``universes''. The pertinent \textsl{LL-brane}
stress-energy tensor \rf{T-S-brane} or, equivalently, \rf{T-S-brane-A}:
\br
T_{(brane)}^{\m\n} = S^{\m\n}\,\d (\eta)
\nonu \\
S^{\m\n} \equiv \frac{T}{\eps\sqrt{b_0}}\,
\llb \pa_\t X^\m \pa_\t X^\n - \eps b_0 G^{ij} \pa_i X^\m \pa_j X^\n 
\rrb_{v=\t,\,\eta=0,\,\th^i =\s^i} 
\lab{T-S-brane-W}
\er
has the following non-zero components with lower indices and trace 
(hereafter we will consider the special case $p=2$ for simplicity):
\be
S_{\eta\eta}= \eps \frac{T}{\sqrt{b_0}} \quad ,\quad
S_{ij} = - T \sqrt{b_0} G_{ij} \quad ,\quad S^\l_\l = - 2 T \sqrt{b_0} \; .
\lab{S-comp-W}
\ee

Let us now turn to the Einstein equations \rf{Einstein-Maxwell-eqs} where 
we explicitly separate the terms contributing to $\d$-function singularities
$\sim \d (\eta)$ on the l.h.s.. These are the terms containing second-order
derivatives w.r.t. $\eta$, since the metric coefficients in
\rf{RN-W-1}--\rf{RN-W-2} are functions of $|\eta|$ and 
$\pa_\eta^2 |\eta| = 2 \d (\eta)$. Thus, we have:
\br
R_{\m\n} \equiv \pa_\eta \G^{\eta}_{\m\n} - \pa_\m \pa_\n \ln \sqrt{-G}
+ \mathrm{non-singular ~terms}
\nonu \\
= 8\pi \Bigl( S_{\m\n} - \h G_{\m\n} S^{\l}_{\l}\Bigr) \d (\eta)
+ \mathrm{non-singular ~terms} \; ,
\lab{E-eqs-sing}
\er
with the \textsl{LL-brane} stress-energy given by \rf{S-comp-W}.
Using the explicit expressions for the Christoffel coefficients:
\be
\G^{\eta}_{vv} = \h {\wti A}\pa_\eta {\wti A} \;\;,\;\;
\G^{\eta}_{v\,\eta} = -\h \pa_\eta {\wti A} \;\; ,\;\; 
\G^{\eta}_{ij} = -\h {\wti A} G_{ij} \pa_\eta \ln {\wti r}^2
\;\; , \;\; \sqrt{-G} = {\wti r}^2
\lab{Christoffel-EF}
\ee
with ${\wti A} (\eta)$ and ${\wti r}(\eta)$ as in \rf{RN-W-2} and 
taking into account ${\wti A}(0) = 0$, it is straightforward to check that non-zero
$\d$-function contributions in $R_{\m\n}$ appear for 
$(\m\n)=(v\,\eta)$ and $(\m\n)=(\eta\eta)$ only. Matching the coefficients
in front of the $\d (\eta)$ on both sides of \rf{E-eqs-sing} yields
accordingly (here for consistency we have to choose $\eps = -1$ in \rf{S-comp-W}):
\be
4\pi T\sqrt{b_0} r^2_{(-)} + r_{(-)} - m = 0 \quad ,\quad
r_{(-)} = \frac{\sqrt{b_0}}{2\pi T} \; ,
\lab{matching-eqs-W}
\ee
with $r_{(-)}$ as in \rf{r-eta} (radius of the inner Reissner-Nordstr{\"o}m
horizon). From \rf{matching-eqs-W} we obtain the following expressions for
the mass and charge parameters of the Reissner-Nordstr{\"o}m ``universes''
as functions of $T$ (the \textsl{LL-brane} dynamical tension) and the free
parameter $b_0$ appearing in the \textsl{LL-brane} dynamics (cf. {\em Appendix}):
\be
m = \frac{\sqrt{b_0}}{2\pi T} (1+2b_0) \quad ,\quad
Q^2 = \frac{b_0}{(2\pi T)^2} (1+4b_0) \; .
\lab{m-Q-eqs}
\ee

The wormhole connecting two non-singular Reissner-Nordstr{\"o}m black hole
``universes'' constructed above is {\em traversable} w.r.t. the {\em proper time}
of in-falling particles (``travelling observers''). This can be directly inferred 
from the equation for point-particle motion (cf. Ref.~\refcite{weinberg}, ch.8, sec.4)
along $\eta$ - ``radial geodesics'':
\be
\eta^{\pr\, 2} + \cV_{\mathrm{eff}}(\eta) = \frac{\cE^2}{m_0^2} \quad ,\quad
\cV_{\mathrm{eff}}(\eta) \equiv {\wti A}(\eta)
\Bigl( 1 + \frac{\cJ^2}{m_0^2 {\wti r}^2(\eta)}\Bigr) \; ,
\lab{particle-eq}
\ee
where the prime indicates proper-time derivative, ${\wti A}(\eta)$ and
${\wti r}(\eta)$ are the same as in \rf{RN-W-2}, and $m_0$, $\cE$ and $\cJ$ 
denote the particle mass, its conserved energy and conserved angular momentum.
The form of the ``effective potential'' $\cV_{\mathrm{eff}}(\eta)$ for
$\cJ=0$ is depicted on Fig.1.

\begin{figure}
\begin{center}
\includegraphics{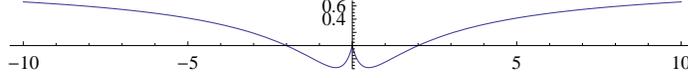}
\caption{Shape of $\cV_{\mathrm{eff}}(\eta)$ as a function of the dimensionless ratio 
$x \equiv \eta/r_{(-)}$}
\end{center}
\end{figure}


However, in accordance with the general arguments \ct{visser-book} 
the \textsl{LL-brane} at the wormhole throat represents an ``exotic matter''
since its stress-energy tensor \rf{T-S-brane-W}--\rf{S-comp-W}
violates the null-energy condition.

\section{Bouncing Cosmology via Lightlike Brane}

Finally, let us consider a extension of the model \rf{E-M-LL} describing
Einstein-Maxwell system interacting with a \textsl{LL-brane} by adding a
{\em bare} cosmological constant term:
\be
S = \int\!\! d^D x\,\sqrt{-G}\,\llb \frac{R(G)}{16\pi} - \frac{\L}{8\pi}
- \frac{1}{4} \cF_{\m\n}\cF^{\m\n}\rrb + S_{\mathrm{LL}} \; .
\lab{E-M-LL+CC}
\ee
Accordingly, the Einstein equations of motion read:
\be
R_{\m\n} - \h G_{\m\n} R  + \L G_{\m\n} =
8\pi \( T^{(EM)}_{\m\n} + T^{(brane)}_{\m\n}\) \; ,
\lab{Einstein-eqs+CC}
\ee
using the same notations as in \rf{E-M-LL}--\rf{Einstein-Maxwell-eqs}.

Eqs.\rf{Einstein-eqs+CC} in the absence of the \textsl{LL-brane}
stress-energy tensor possess spherically symmetric Reissner-Nordstr{\"o}m-de-Sitter
solution (here again we consider $D=4$):
\be
ds^2 = - A_\L (r) dt^2 + \frac{dr^2}{A_\L (r)} + r^2 \(d\th^2 + \sin^2\th d\vp^2\) \; ,
\lab{RN+dS-0}
\ee
or, written in Eddington-Finkelstein coordinates ($dt=dv - \frac{dr}{A_\L (r)}$):
\br
ds^2 = - A_\L (r) dv^2 + 2 dv\,dr + r^2 \(d\th^2 + \sin^2\th d\vp^2 \)  \; ,
\nonu \\
A_\L (r) = 1 - \frac{2m}{r} + \frac{Q^2}{r^2} - \frac{\L}{3} r^2 \; .
\lab{RN+dS}
\er
For sufficiently large $\L$ there is only one {\em single} horizon at 
$r= r_{(0)}\equiv r_{(0)}(m,Q^2,\L)$, where:
\be
A_\L (r_{(0)}) = 0 \quad ,\quad \pa_r A_\L \bv_{r=r_{(0)}} < 0 \quad ,
\;\; \mathrm{i.e.}\;\; A_\L (r) <0 \;\; \mathrm{for}\;\; r>r_{(0)} \; .
\lab{RN+dS-hor}
\ee

Let us now consider the following modification of the Reissner-Nordstr{\"o}m-de-Sitter
metric \rf{RN+dS}:
\br
ds^2 = {\wti A}_\L (\eta) dv^2 + 2 dv\,d\eta + 
{\wti r}^2 (\eta) \(d\th^2 + \sin^2\th d\vp^2 \)  \; ,
\lab{RN+dS-1} \\
{\wti A}_\L (\eta) \equiv - A_\L (r_{(0)} + |\eta|) = \frac{\L}{3} {\wti r}^2 (\eta) - 
\frac{Q^2}{{\wti r}^2 (\eta)} + \frac{2m}{{\wti r} (\eta)} -1 \; ,
\lab{RN+dS-2}
\er
where ${\wti r} (\eta) = r_{(0)} + |\eta|$ and as above $\eta$ varies from 
$-\infty$ to $+\infty$. Because of \rf{RN+dS-hor} the metric component 
${\wti A}_\L$ \rf{RN+dS-2} in \rf{RN+dS-1} is always non-negative:
\be
{\wti A}_\L (\eta) > 0 \;\; \mathrm{for}\;\; \eta >0 \; \mathrm{or}\;
\eta <0 \quad ;\quad {\wti A}_\L (0) = 0 \; .
\lab{A-L}
\ee
The metric \rf{RN+dS-1}--\rf{RN+dS-2} corresponds to a space-time manifold
consisting of two identical copies of the {\em exterior} Reissner-Nordstr{\"o}m-de-Sitter
space-time region ($r>r_{(0)}$) -- one ``universe'' for $\eta >0$ and another 
identical ``universe'' for $\eta <0$, 
glued together along the single Reissner-Nordstr{\"o}m-de-Sitter horizon  at 
$r=r_{(0)}$, \textsl{i.e.}, $\eta =0$. Since in the exterior
Reissner-Nordstr{\"o}m-de-Sitter region the coordinates $(t,r)$ exchange
their roles, whereupon $r$ becomes timelike, the same is true for the coordinate $\eta$
in \rf{RN+dS-1}--\rf{RN+dS-2}, \textsl{i.e.}, $\eta$ is timelike coordinate
in both ``universes'' except at the matching hypersurface $\eta=0$. 
The latter is directly seen upon transforming \rf{RN+dS-1} into diagonal form 
($(v,\eta) \to (\xi,\eta)$ with $d\xi = dv - \frac{d\eta}{{\wti A}_\L (\eta)}$):
\be
ds^2 = - \frac{d\eta^2}{\wti{A}_\L (\eta)} + \wti{A}_\L (\eta) d\xi^2 +
{\wti r}^2 (\eta) \(d\th^2 + \sin^2\th d\vp^2 \)  
\lab{RN+dS-1-diag}
\ee
with $\wti{A}_\L (\eta)$ as in \rf{RN+dS-2} and $\eta \in (-\infty,+\infty)$.

Now, repeating the same steps as in the derivation of the wormhole connecting two
non-singular black holes in the first part of the present Section 
(Eqs.\rf{RN-W-1}--\rf{m-Q-eqs}) we obtain that the metric 
\rf{RN+dS-1}--\rf{RN+dS-2} is a solution of the Einstein equations \rf{Einstein-eqs+CC}
with the \textsl{LL-brane} stress-energy tensor {\em included}. Here again 
$T^{(brane)}_{\m\n}$ has the same form as in \rf{T-S-brane-W}--\rf{S-comp-W}, whereas the
matching of the coefficients in front of the delta-functions $\d (\eta)$ on both sides
of \rf{Einstein-eqs+CC} yield (cf. Eqs.\rf{matching-eqs-W} above):
\be
\frac{\L}{3} r_{(0)} + \frac{Q^2}{r_{(0)}^3} - \frac{m}{r_{(0)}^2}= 4\pi T \sqrt{b_0}
\quad ,\quad
r_{(0)} = \frac{\sqrt{b_0}}{2\pi T} \; ,
\lab{matching-eqs-CC}
\ee
where $r_{(0)}\equiv r_{(0)}(m,Q^2,\L)$ is the single Reissner-Nordstr{\"o}m-de-Sitter
horizon radius \rf{RN+dS-hor}. Eqs.\rf{matching-eqs-CC} determine $m$ and $Q^2$
as functions of the bare cosmological constant $\L$, and of $T$ (the \textsl{LL-brane}
dynamical tension) and the free parameter $b_0$ of the \textsl{LL-brane} dynamics 
(cf. {\em Appendix}).

The physical meaning of the space-time manifold with the metric 
\rf{RN+dS-1}--\rf{RN+dS-2} or \rf{RN+dS-1-diag} can be more clearly seen upon making the 
following change of the time coordinate $\eta \to \bar{\eta}$ with:
\be
d\bar{\eta} = \frac{d\eta}{\sqrt{{\wti A}_\L (\eta)}} \quad ,\quad
\bar{\eta} \simeq \frac{1}{2 \sqrt{A_0}} \mathrm{sign} (\eta) \sqrt{|\eta|}
\;\;\; \mathrm{for ~small}\;\;\eta \;\; ,
\lab{eta-bar}
\ee
where $A_0 \equiv -\pa_r A_\L\!\bv_{r=r_{(0)}}>0$, such that \rf{RN+dS-1-diag} becomes:
\br
ds^2 = - d\bar{\eta}^2 + \widehat{A}_\L (\bar{\eta}) d\xi^2 +
\widehat{r}^2 (\bar{\eta}) \(d\th^2 + \sin^2\th d\vp^2 \) \; ,
\lab{RN+dS-3} \\
\widehat{A}_\L (\bar{\eta}) = {\wti A}_\L \(\eta (\bar{\eta})\) =
- A_\L (r_{(0)} + |\eta (\bar{\eta})|) \quad ,\quad
\widehat{r} (\bar{\eta}) = r_{(0)} + |\eta (\bar{\eta})| \; ,
\lab{RN+dS-4}
\er
(here $A_\L$ and $r_{(0)}$ are the same as in \rf{RN+dS}--\rf{RN+dS-hor}). The 
coordinate singularity of the metric \rf{RN+dS-1-diag} at $\eta=0$ or of the metric 
\rf{RN+dS-3} at $\bar{\eta}=0$ is purely artificial one as it is absent in the
Eddington-Finkelstein form \rf{RN+dS-1}.

The new timelike coordinate $\bar{\eta}$ varies again from $-\infty$ to $+\infty$.
The metric \rf{RN+dS-3} is a special case of the anisotropic Kantowski-Sachs cosmology 
\ct{KS-cosmolog}, where the three spatial dimensions are having a dynamical evolution.
In the present case the two spherical angular dimensions are having the same
dynamical evolution, which produces a bounce from a non-zero value, while the 
additional spatial dimension $\xi$ exhibits a different
bounce behavior, namely bouncing from zero size although this is not an
intrinsic space-time singularity as explained above.
A distinguishing feature of the present solution is that the universe evolving
from negative time ($\bar{\eta}<0$) undergoes contraction until
$\bar{\eta}=0$, but then proceeds for $\bar{\eta}>0$ with an expansion instead of 
contraction.

\section{Discussion and Conclusions}

In the present paper we have explicitly constructed a non-singular black hole
as a self-consistent solution of the equations of motion derived from a
well-defined lagrangian action principle for a bulk Einstein-Maxwell-Kalb-Ramond 
system coupled to a codimension-one charged {\em lightlike} $p$-brane with 
dynamical (variable) tension (\textsl{LL-brane}). We stress on the fact that
the latter is described by a manifestly reparametrization-invariant world-volume 
action of either Polyakov-type or Nambu-Goto-type which is significantly different 
from the ordinary Nambu-Goto $p$-brane action. The crucial point in our
construction is that \textsl{LL-brane} turns out to be the proper
gravitational and charge source 
in the Einstein-Maxwell-Kalb-Ramond equations of motion needed to generate a 
self-consistent solution describing {\em non-singular} black hole. 
The latter consists of de
Sitter interior region and exterior Reissner-Nordstr{\"o}m region glued
together along their common horizon, which is the inner horizon from the 
Reissner-Nordstr{\"o}m side. The matching horizon is automatically occupied 
by the \textsl{LL-brane} as a result of its world-volume lagrangian dynamics, 
which also generates the cosmological constant in the interior region and
uniquely determines the mass and charge parameters of the exterior region.
In particular, Eq.\rf{metric-matching-1} tells us that the size, mass and
charge of the non-singular black hole might be very small for large $\b$,
\textsl{i.e.}, provided the \textsl{LL-brane} is strongly coupled to the
bulk Kalb-Ramond gauge field.

Further, we have constructed a self-consistent {\em wormhole} solution of
Einstein-Maxwell system coupled to electrically neutral \textsl{LL-brane},
which describes two regular black hole space-time regions matched along the
``throat'' which is their common horizon. The two black hole ``universes''
are identical copies of the exterior Reissner-Nordstr{\"o}m region above the inner 
Reissner-Nordstr{\"o}m horizon -- the ``throat'', which is now 
occupied by the \textsl{LL-brane}. The corresponding mass and charge
parameters of the black hole ``universes'' are explicitly determined 
by the dynamical \textsl{LL-brane} tension. This also provides an explicit
example of Misner-Wheeler ``charge without charge'' phenomenon.
Provided a sufficiently large bare cosmological constant is added, we have
shown that the above wormhole solution connecting two non-singular black
holes can be transformed into a 
cosmological bouncing solution of Kantowski-Sachs type.


Mechanisms for singularity avoidance in regular black hole solutions have
been thoroughly analyzed in Refs.~\refcite{borde}. Similarly to the previously
derived non-singular black holes, the existence and consistency of the
presently proposed regular black hole solution (Section 5) can be attributed
to the topology change in the structure of the corresponding spacelike slices.

At this point the issue of stability remains an open question. Let us note
that the latter is greatly affected by the nature of the matching shell
dynamics, see \textsl{e.g.} Ref.~\refcite{BGG}. It was shown there that if the
{\em timelike} shell separating a Schwarzschild region from a de Sitter region obeys a
``domain wall'' equation of state, then the equation of motion of the wall
corresponds to the equation of motion of a particle in a potential with only
one stationary point which is a maximum. This does not tell us a lot about
the \textsl{LL-brane} case since the latter are not ordinary domain walls.
In fact, when additional matter is added on the wall, the wall potential may
acquire a minimum \ct{guendel-portnoy}. In addition to this, in case when
the surface tension becomes dynamically zero at certain shell radius, the potential of 
the shell displays an attractive behaviour towards that same radius 
\ct{guendel-sakai}. Another argument in favor of the stability of the solution
in the case of Reissner-Nordstr{\"o}m exterior region and equality of the
radial pressures on both sides of the shell (\textsl{i.e.}, zero surface tension)
is the following fact. If the radius of the shell is lowered, the
de Sitter pressure becomes bigger than the Reissner-Nordstr{\"o}m pressure.
The opposite takes place when the shell radius is increased. All this indicates a
``mechanical'' argument for stability. There is, however, a ``kinematical''
argument against stability since a perturbation that makes a shell radius
bigger than the de Sitter horizon will force the shell to expand to infinity
leading to the creation of a child universe \ct{BGG}.

It is not clear how the above arguments can be applied to the case of
\textsl{LL-branes} since the \textsl{LL-brane} world-volume dynamics forces
the latter to locate itself automatically on certain horizon of the
embedding space-time metrics.

The regular black hole solution via \textsl{LL-brane} constructed above
may still be interesting even if it turns out to be unstable, since the
instability (if it exists) appears to be towards formation of a child universe.
Furthermore, unstable solutions may play an important role in quantum effects
like sphaleron solutions \ct{klink} in Glashow-Weinberg-Salam model. All these 
issues should be a subject of further studies.

Finally, let us mention that quantum effects may have significant impact on the fate of 
black hole singularities, namely, removing them altogether 
\ct{trieste-ncg,stojkovic-etal-quantum}.
It would be very interesting to study the quantization of the above
Einstein-Maxwell-Kalb-Ramond system interacting with charged \textsl{LL-brane} where
black hole singularities have been removed already at classical level by the
presence of the \textsl{LL-brane} alone.

\appendix

\section{Nambu-Goto-Type World-Volume Formulation of Lightlike Branes}

Lets us consider the {\em dual} Nambu-Goto-type world-volume action of the
\textsl{LL-brane}:
\be
S_{\rm NG} = - \int d^{p+1}\s \, T 
\sqrt{\bgv\, \det\Vert g_{ab} - \eps \frac{1}{T^2}\pa_a u \pa_b u\Vert\,\bgv}
\quad ,\quad \eps = \pm 1 \; ,
\lab{LL-action-NG-A}
\ee
where $g_{ab}$ indicates the induced metric on the world-volume \rf{ind-metric}.
The choice of the sign in \rf{LL-action-NG-A} does not have physical effect
because of the non-dynamical nature of the $u$-field. 

The corresponding equations of motion w.r.t. $X^\m$, $u$ and $T$ read:
\br
\pa_a \( T \sqrt{|{\wti g}|} {\wti g}^{ab}\pa_b X^\m\)
+ T \sqrt{|{\wti g}|} {\wti g}^{ab} \pa_a X^\l \pa_b X^\n \G^\m_{\l\n} = 0 \; ,
\lab{X-eqs-NG}\\
\pa_a \(\frac{1}{T} \sqrt{|{\wti g}|} {\wti g}^{ab}\pa_b u\) = 0 \; ,
\lab{u-eqs-NG} \\
T^2 + \eps {\wti g}^{ab}\pa_a u \pa_b u = 0 \; ,
\lab{T-eq-NG}
\er
where we have introduced the convenient notations:
\be
{\wti g}_{ab} = g_{ab} - \eps \frac{1}{T^2}\pa_a u \pa_b u \quad ,\quad
{\wti g} \equiv \det\Vert {\wti g}_{ab}\Vert \; ,
\lab{ind-metric-ext}
\ee
and ${\wti g}^{ab}$ is the inverse matrix w.r.t. ${\wti g}_{ab}$.

From the definition \rf{ind-metric-ext} and Eq.\rf{T-eq-NG} one easily finds
that the induced metric on the world-volume is singular on-shell (cf. 
Eq.\rf{on-shell-singular} above):
\be
g_{ab} \( {\wti g}^{bc}\pa_c u\) = 0
\lab{on-shell-singular-A}
\ee
exhibiting the lightlike nature of the $p$-brane described by \rf{LL-action-NG-A}.

Similarly to the treatment of the \textsl{LL-brane} dynamics within the 
Polyakov-type formulation (Section 3 above) we can choose the following
gauge-fixing of world-volume reparametrization invariance (cf. Eqs.\rf{gauge-fix}):
\be
{\wti g}_{0i} = 0 \;\; (i=1,\ldots,p) \quad ,\quad 
{\wti g}_{00} = -\eps b_0 \; ,\; b_0 = \mathrm{const} >0 \; .
\lab{gauge-fix-A}
\ee
Also, we will use a natural ansatz :
\be
u = u (\t) \;\; ,\; \mathrm{i.e.}, \;\; \pa_i u = 0 \; ,
\lab{u-ansatz}
\ee
which means that we choose the lightlike direction in Eq.\rf{on-shell-singular-A} 
to coincide with the brane proper-time direction on the world-volume (this
is analogous to the ansatz \rf{F-ansatz} within the Polyakov-type formulation).
With \rf{gauge-fix-A}--\rf{u-ansatz} Eq.\rf{u-eqs-NG} implies (cf. 
Eq.\rf{gamma-p-0} above):
\be
\pa_0 g^{(p)} = 0 \quad \mathrm{where}\;\; g^{(p)}\equiv\det\Vert g_{ij}\Vert \; ,
\lab{g-p-0}
\ee
with $g_{ij}$ -- the spacelike part of the induced metric \rf{ind-metric}.

Taking into account \rf{gauge-fix-A}--\rf{g-p-0}, the equations of motion
\rf{X-eqs-NG},\rf{u-eqs-NG} and \rf{T-eq-NG} (or, equivalently, 
\rf{on-shell-singular-A}) reduce to (cf. Eqs.\rf{gamma-eqs-0}--\rf{X-eqs-0}):
\br
g_{00} \equiv \Xdot^\m\!\! G_{\m\n}\!\! \Xdot^\n = 0 \quad ,\quad 
g_{0i} \equiv \Xdot^\m\!\! G_{\m\n} \pa_i X^\n = 0 \quad ,\quad
\pa_i T = 0 \; ,
\lab{g-eqs-A} \\
-\pa_0 \( T \pa_0 X^\m\) - \frac{T \eps b_0}{\sqrt{g^{(p)}}}
\pa_i \(\sqrt{g^{(p)}} g^{ij} \pa_j X^\m\)
\nonu \\
+ T \(\pa_0 X^\n \pa_0 X^\l - \eps b_0 g^{kl} \pa_k X^\n \pa_l X^\l\)
\G^\m_{\n\l} = 0 \; .
\lab{X-eqs-A}
\er
The \textsl{LL-brane} stress-energy tensor derived from the
Nambu-Goto-type action \rf{LL-action-NG-A} reads (cf.\rf{T-brane}):
\be
T_{(brane)}^{\m\n} = 
- \int\!\! d^{p+1}\s\,\frac{\d^{(D)}\Bigl(x-X(\s)\Bigr)}{\sqrt{-G}}\,
T\,\sqrt{|{\wti g}|} {\wti g}^{ab}\pa_a X^\m \pa_b X^\n \; ,
\lab{T-brane-A}
\ee
where the short-hand notation \rf{ind-metric-ext} is used. For the 
embedding \rf{X-embed} and taking into account \rf{gauge-fix-A} we obtain:
\br
T_{(brane)}^{\m\n} = S^{\m\n}\,\d (r-r_0)
\nonu \\
S^{\m\n} \equiv \frac{T}{\eps\sqrt{b_0}}\,
\llb \pa_\t X^\m \pa_\t X^\n - \eps b_0 G^{ij} \pa_i X^\m \pa_j X^\n 
\rrb_{v=\t,\,r=r_0,\,\th^i =\s^i} \; ,
\lab{T-S-brane-A}
\er
where $r_0$ is a horizon of the bulk space-time metric.
The \textsl{LL-brane} stress-energy tensor \rf{T-S-brane-A} is identical to the 
one obtained within the Polyakov-type formulation (Eq.\rf{T-S-brane} above)
with the identifications:
\be
\chi = T b_0^{\frac{p-1}{2}} \eps^{\frac{p-2}{2}} \quad ,\quad
2a_0 = \eps b_0 \; ,
\lab{Polya-Nambu-comp}
\ee
where $\chi$, $a_0$ and $b_0$ are the same as in \rf{LL-action-chi}, 
\rf{a0-const} and \rf{gauge-fix-A}, respectively.

\section*{Acknowledgments}
E.N. and S.P. are supported by Bulgarian NSF grant \textsl{DO 02-257}.
Also, all of us acknowledge support of our collaboration through the exchange
agreement between the Ben-Gurion University of the Negev (Beer-Sheva, Israel) and
the Bulgarian Academy of Sciences.


\end{document}